\documentclass [11pt,twoside]{article}
\usepackage[cp1251]{inputenc}
\usepackage[english,ukrainian]{babel}
\usepackage{amssymb}\usepackage{amsmath}
\usepackage[pdfstartview=FitH,pdfnewwindow=true,colorlinks,pagebackref,breaklinks,bookmarks=false]{hyperref}
\addto\captionsukrainian{}
\addto\captionsukrainian{}
\Ukrainian
\newcommand{\updot}{\displaystyle\boldsymbol\cdot}
\begin{document}
\title{\large\bf ВАРІЯЦІЙНЕ УЗАГАЛЬНЕННЯ\\ВІЛЬНОЇ РЕЛЯТИВІСЬКОЇ
ДЗИҐИ}
\author{\href{http://www.iapmm.lviv.ua/12/files/st_files/matsyuk.htm}{\em Роман МАЦЮК}}
\date{Інститут прикладних проблем механіки і математики \\ ім.~Я.~С.~Підстригача НАН~України,\\
вул.~Наукова,~3б, Львів~79060\\
e-mail: matsyuk@lms.lviv.ua, romko.b.m@gmail.com}
 \maketitle
\thispagestyle{myheadings} \markboth{~\hrulefill~Фізичний збірник
НТШ т.",6 2006 p.}{Фізичний збірник НТШ т.6 2006 p.~\hrulefill~}
{\hfill \small Редакція отримала статтю 11 листопада 2005 р.}
\renewcommand{\thefootnote}{}\footnotetext{PACS numbers
04.25.-g, 03.30.+p, 03.65.Sq, 45.10.Db}
\begin{abstract}
\hspace*{-15pt}Показуємо, що добре знані рівняння руху першого
порядку (за спіном, імпульсом та координатою) для релятивіської
дзиґи є рівнозначними до рівнянь третього порядку Матісона усюди
на поверхні додаткової в'язі Матісона-Пірані. Опісля розглядаємо
ці рівняння третього порядку в пласк\'ому просторі-часі та й
винаходимо функцію Ляґранжа для них. Коли дозволити фізичну
інтерпретацію усієї множини екстремалей, то як наслідок відкриємо
цілий спектр власної маси релятивіської дзиґи, параметризований
значеннями власного моменту.
\end{abstract}

\section{Вступ}
Поведінку крутької буцім-клясичної частки, наділеної скісним
тензором $S^{ij} = S^{{\left[ {ij} \right]}}$, можна описати
відомою системою рівнянь (гляди,
на\-п\-ри\-к\-лад,~\cite{matsyuk:3})
\begin{gather}\label{matsyuk:1}
\dot {P}^{i} = -{} \dfrac{1}{2} R^{i}_{jkl} u^{j}S^{kl}
 \\
\dot {S}^{ij} = P^{i}u^{j} - P^{j}u^{i}{\rm .}
\label{matsyuk:2}
\end{gather}

Цих рівнянь недосить для знаходження світової лінії частки. Тому їх
доповнюють ріжними додатковими умовами. Часто послуговуються умовою
Тульчиєва
\begin{equation}\label{matsyuk:3}
P_{j} S^{ij} = 0{\rm .}
\end{equation}

Ми ж використовуватимемо умову Матісона-Пірані
\begin{equation}\label{matsyuk:4}
u_{j} S^{ij} = 0{\rm .}
\end{equation}

Ріжницю між умовами (\ref{matsyuk:3}) і (\ref{matsyuk:4}) трактуємо як ріжницю між способами
встановлення залежности поміж змінними ${\rm {\bf P}}$ та ${\rm {\bf u}}$.
Перехід від швидкости до імпульсу можна оцінювати як впровадження фазового
простору. У варіяційному численні такий перехід називають (узагальненим)
перетворенням Лєжандра. Ми виконаємо цю проґраму в пласк\'ому просторі-часі.
Додаткова умова Матісона-Пірані дозволить доволі швидко перейти до
варіяційної форми рівняння світової лінії частки. Відповідна функція
Ляґранжа задасть шукане перетворення Лєжандра. Правда, на цій дорозі
підвищиться порядок похідних у варіяційному рівнянні. Таким чином, йде мова
про формалізм механіки Остроградського.

\markboth{~\hrulefill~Варіяційне узагальнення вільної...}{Р.",Мацюк~\hrulefill~}

Факт підвищення порядку похідних у рівнянні, яке описує динаміку просторових
координат релятивіської дзиґи, зістав наголошений ще у 1945\textsuperscript{му}~році
Вайсенгофом~\cite{matsyuk:6} у його поклику на статтю Матісона~\cite{matsyuk:4}.
Матісон,
додатково послуговуючись в'яззю (\ref{matsyuk:4}), отримав для світової лінії дзиґи в
ґравітаційному полі таке рівняння:
\begin{equation}\label{matsyuk:5}
m\dot {u}^{i} = \ddot {u}_{j} S^{ij} - {\frac{{1}}{{2}}}R^{i}_{jkl}
u^{j}S^{kl}{\rm .}
\end{equation}

Побутує думка, що рівняння (\ref{matsyuk:5}) не є рівнозначне до системи рівнянь (\ref{matsyuk:1},~\ref{matsyuk:2}).
Ми покажемо, що рівнозначність зберігається усюди на поверхні (4).

Повертаючись до вільної дзиґи, ми запропонуємо варіяційне \textit{узагальнення} рівняння третього
порядку (\ref{matsyuk:5}), яке дозволить „розморозити“ значення „$0$“ інтеґрала руху
(в пласк\'ому просторі)
\begin{equation}\label{matsyuk:6}
{\frac{{{\rm {\bf   {{s}}} } \cdot {\rm {\bf u}}}}{{{\left\| {{\rm {\bf u}}}
\right\|}}}}{\rm ,}
\end{equation}
де значком ${\rm {\bf   {{s}}} }$ позначений деякий чотири-вектор „спіну“
(вертуна). Тільки в цьому „розмороженні“ і полягатиме нерівнозначність між
\textit{узагальненим} рівнянням руху вільної дзиґи (\ref{matsyuk:26}) і системою (1,~2,~4) без ґравітації. Наша
тактика полягає в тім, щоби, отримавши в пласк\'ому просторі \textit{варіяційне} рівняння третього порядку (\ref{matsyuk:26}), яке
містить усі розв'язки системи рівнянь (5,~2,~4) при $R^{i}_{jkl}=0$, і яке ми власне називаємо
\textit{узагальненим,} відважитись розглядати і всі решта розв'язків такого \textit{узагальненого} рівняння.
Виявляється, що ця решта розв'язків \textit{не виходить за межі неперервного спектру інтеґрала руху} (\ref{matsyuk:6}). Цієї властивості не добитись,
просто відкидаючи умову (\ref{matsyuk:4}) і залишаючись при рівнянні (\ref{matsyuk:5}). Можна сказати,
що „розмороження“ інтеґрала (\ref{matsyuk:6}) є рівнозначним до варіяційної модифікації
рівняння (\ref{matsyuk:5}). При цьому виявляється, що \textit{уся множина розв'язків узагальненого рівняння складається з таких і тільки з таких світових
ліній, яких можна інтерпретувати як рухи крутьких часток з деякою позірною масою}
\begin{equation}\label{matsyuk:7}
m = \mu \left( {1 - {\frac{{\left( {{\rm {\bf   {{s}}} } \cdot {\rm {\bf u}}}
\right)^{2}}}{{{\rm {\bf   {{s}}} }^{2}{\rm {\bf u}}^{2}}}}}
\right)^{{\raise0.7ex\hbox{${3}$} \!\mathord{\left/ {\vphantom {{3}
{2}}}\right.\kern-\nulldelimiterspace}\!\lower0.7ex\hbox{${2}$}}}{\rm .}
\end{equation}
Частки з масою $m \ne \mu $ рухаються так, щоби чотири-вектор „спіну“ ${\rm
{\bf   {{s}}} }$, завдяки інтеґралові руху (\ref{matsyuk:6}), утворював постійний кут з напрямом чотири-швидкости.

\section{Рівняння Матісона}\label{sec:mylabel1}
Розглянемо деякі рівнозначні перетворення системи рівнянь (1,~2,~4).

\textbf{Твердження.}\textbf{\textit{} }\textit{В області} ${\rm {\bf u}}^{2} \ne 0$
\textit{рівняння (\ref{matsyuk:2}) алґебрично рівнозначне таким двом:}
\begin{gather}\label{matsyuk:8}
{\rm {\bf u}}^{2}P^{i} - \left( {{\rm {\bf P}} \cdot {\rm {\bf u}}}
\right)\;u^{i} = u_{j} \dot {S}^{ij}
 \\
\varepsilon _{ijkl} u^{j}\dot {S}^{kl} = 0{\rm .}
\label{matsyuk:9}
\end{gather}

\textbf{\textit{Доведення}}. Щоби переконатись в імплікації
(\ref{matsyuk:8},~\ref{matsyuk:9})
$ \Rightarrow $ (\ref{matsyuk:2}), потрібно домножити рівняння
(\ref{matsyuk:9}) зовнішнім чином
справа на вектор ${\rm {\bf u}}$, перейти до двоїстого рівняння і
відняти від цього подвоєний зовнішній добуток рівняння (\ref{matsyuk:8}) на
вектор ${\rm {\bf u}}$ (комбінація
$\frac12\,e^{ijkl}u_{l} \left( {9} \right)_{k} - 2u^{{\left[ {j}
\right.}}\left( {8} \right)^{{\left. {i} \right]}}$ дає рівняння
$\left( {2} \right)^{ij}$, де $e^{ijkl} = \delta
_{0}^{i}{}_{1}^{j} {}_{2}^{k} {}_{3}^{l} $ -- одиничний повністю
скісний (антисиметричний) контраваріянтний псевдотензор Леві-Чивіти).~$\blacksquare$

Виходить, замість системи рівнянь (\ref{matsyuk:1},~\ref{matsyuk:2}) можна розглядати
систему (\ref{matsyuk:1},~\ref{matsyuk:8},~\ref{matsyuk:9}).

Пустимо тепер в хід в'язь (\ref{matsyuk:4}) разом з її диференційним продовженням,
\begin{equation}\label{matsyuk:10}
u_{j} \dot {S}^{ij} = S^{ji}\dot {u}_{j} {\rm .}
\end{equation}
С\'аме рівність (\ref{matsyuk:10}) дозволяє запровадити інтеґрал руху
\begin{equation}\label{matsyuk:11}
m = {\frac{{{\rm {\bf P}} \cdot {\rm {\bf u}}}}{{{\left\| {{\rm {\bf u}}}
\right\|}}}}{\rm ,}
\end{equation}
оскільки алґебричним наслідком рівнянь
(\ref{matsyuk:1},~\ref{matsyuk:8},~\ref{matsyuk:10}) є наступна рівність, яка у
своїй лівій частині пропорційна до упохідненої правої частини виразу (\ref{matsyuk:11}):
\begin{equation}\label{matsyuk:12}
{\rm {\bf u}}^{2}\left( {{\rm {\bf P}} \cdot {\rm {\bf u}}} \right)^{^{
\updot} } - {\frac{{1}}{{2}}}\left( {{\rm {\bf P}} \cdot {\rm {\bf u}}}
\right)\left( {{\rm {\bf u}}^{2}} \right)^{^{\updot} } = 0{\rm .}
\end{equation}

Диференційна умова Матісона-Пірані (\ref{matsyuk:10}) є вирішальною для можливости
повністю усунути члени, які містять похідну від тензора спіну і неявно є
присутні у лівій частині рівняння (\ref{matsyuk:1}). З урахуванням умови
(\ref{matsyuk:10}) означення
імпульсу (\ref{matsyuk:8}) прибере форми:
\begin{equation}\label{matsyuk:13}
{\rm {\bf u}}^{2}P^{i} - \left( {{\rm {\bf P}} \cdot {\rm {\bf u}}}
\right)\;u^{i} = \dot {u}_{j} S^{ji}{\rm ,}
\end{equation}
з її алґебричним наслідком
\begin{equation}\label{matsyuk:14}
{\rm {\bf u}}^{2}\left( {{\rm {\bf P}} \cdot {\rm {\bf \dot {u}}}} \right) =
\left( {{\rm {\bf P}} \cdot {\rm {\bf u}}} \right)\,\left( {{\rm {\bf u}}
\cdot {\rm {\bf \dot {u}}}} \right){\rm .}
\end{equation}
Справедливим, отже, є таке

\textbf{Твердження.} \textit{На многовиді (\ref{matsyuk:10})
при}
${\rm {\bf u}}^{2} \ne 0$\textit{ рівняння (\ref{matsyuk:2}) алґебрично
рівнозначне як зі системою рівнянь (\ref{matsyuk:8},~\ref{matsyuk:9}),
так і зі
системою рівнянь (\ref{matsyuk:13},~\ref{matsyuk:9}).}

Наступним кроком вилучім змінну величину ${\rm {\bf P}}$ із системи рівнянь
(\ref{matsyuk:1},~\ref{matsyuk:10}, \ref{matsyuk:9},~\ref{matsyuk:13}).
З цією метою вдамося до першого диференційного продовження
рівняння (\ref{matsyuk:13}):
\begin{equation}\label{matsyuk:15}
P^{i}\left( {{\rm {\bf u}}^{2}} \right)^{^{\updot} } + {\rm {\bf u}}^{2}\dot
{P}^{i} - \left( {{\rm {\bf P}} \cdot {\rm {\bf u}}} \right)\;\dot {u}^{i} -
u^{i}\;\left( {{\rm {\bf P}} \cdot {\rm {\bf u}}} \right)^{^{\updot} } =
S^{ji}\ddot {u}_{j} + \dot {u}_{j} \dot {S}^{ji}{\rm .}
\end{equation}
Замість рівняння (\ref{matsyuk:1}) розгляньмо таке:
\begin{equation}\label{matsyuk:16}
{\frac{{1}}{{2}}}\left( {{\rm {\bf u}}^{2}} \right)^{^{\updot} }{\left[
{\left( {{\rm {\bf P}} \cdot {\rm {\bf u}}} \right)\;u^{i} - 3\,S^{ij}\dot
{u}_{j}}  \right]} - {\rm {\bf u}}^{2}\left( {{\rm {\bf P}} \cdot {\rm {\bf
u}}} \right)\;\dot {u}^{i} + {\rm {\bf u}}^{2}S^{ij}\ddot {u}_{j} =
{\frac{{1}}{{2}}}\left( {{\rm {\bf u}}^{2}} \right)^{2}R^{i}_{jkl}
u^{j}S^{kl}{\rm .}
\end{equation}

\textbf{Твердження}\textit{. В області} ${\rm {\bf u}}^{2} \ne 0$\textit{ на
многовиді (\ref{matsyuk:10}) система рівнянь
(\ref{matsyuk:1},~\ref{matsyuk:2},~\ref{matsyuk:15}) алґебрично рівнозначна
зі системою (\ref{matsyuk:16},~\ref{matsyuk:9},~\ref{matsyuk:13},
\ref{matsyuk:15}, \ref{matsyuk:12}).}

\textbf{\textit{Доведення}}. Щоби переконатись у справедливості імплікації
(\ref{matsyuk:1},~\ref{matsyuk:2},~\ref{matsyuk:15})
$ \Rightarrow $ (\ref{matsyuk:16},~\ref{matsyuk:9},
\ref{matsyuk:13}, \ref{matsyuk:15}, \ref{matsyuk:12}),
підставимо вираз для похідної
$\dot {S}^{ij}$ з рівняння (\ref{matsyuk:2}) у рівняння (\ref{matsyuk:15})
і домножимо усе на ${\rm {\bf u}}^{2}$.
Далі нагадаємо, що алґебричним наслідком системи
(\ref{matsyuk:1},~\ref{matsyuk:8},~\ref{matsyuk:10}) є рівняння
(\ref{matsyuk:12}) і (\ref{matsyuk:13}). Тому використаємо (\ref{matsyuk:12}) і, замінивши
вираз ${\rm {\bf u}}^{2}{\rm {\bf P}}$ з рівняння (\ref{matsyuk:13}),
а вираз для ${\rm {\bf \dot {P}}}$ з рівняння
(\ref{matsyuk:1}), одержимо рівняння (\ref{matsyuk:16}).

Щоби переконатися у справедливості зворотньої імплікації, від рівняння
(\ref{matsyuk:16})
віднімемо рівняння (\ref{matsyuk:15}) з коефіцієнтом ${\rm {\bf u}}^{2}$.
За допомоги
співвідношення (\ref{matsyuk:10}) вилучимо доданок
$ - 3\,\dot {u}_{j} S^{ij}$, а опісля,
ґрунтуючись на попередньому Твердженні, замінимо похідну $\dot {S}^{ij}$ з
рівняння (\ref{matsyuk:2}). Далі скористаємо з наслідка (\ref{matsyuk:14}).
Врешті, застосувавши (\ref{matsyuk:12}),
відразу отримаємо рівняння (\ref{matsyuk:1}).~$\blacksquare$

Зауважимо, що до цього часу ми ще не користали з додаткової умови
Матісона-Пірані (\ref{matsyuk:4}), а тільки з її диференційного
продовження (\ref{matsyuk:10}). Так
само, як і не вибирали іще конкретного значення інтеґрала руху
${\frac{{{\rm {\bf P}} \cdot {\rm {\bf u}}}}{{{\left\| {{\rm {\bf u}}}
\right\|}}}}$
формулою
(\ref{matsyuk:11}).

Коли привести рівняння (\ref{matsyuk:16}) до натурального параметра та
закріпити значення
величини маси формулою (\ref{matsyuk:11}), тоді воно співпаде з рівнянням
третього порядку
(\ref{matsyuk:5}) для поступального руху буцім-клясичної крутької частки,
яке можна
вичитати зі статті Матісона~\cite{matsyuk:4}.

В системі рівнянь (\ref{matsyuk:16},~\ref{matsyuk:9},
\ref{matsyuk:13}, \ref{matsyuk:12}) формула (\ref{matsyuk:13})
 просто дає означення
імпульса, тому її можна опустити з точки зору пошуку розв'язків.

Кількість рівнянь у системі
(\ref{matsyuk:16},~\ref{matsyuk:9}, \ref{matsyuk:4},
\ref{matsyuk:10}) можна значно зменшити,
запровадивши чотири-векторну величину „вертуна“ (або буцім-спіна)
\begin{equation}\label{matsyuk:17}
  {{s}} _{i} = {\frac{{\sqrt {{\left| {g} \right|}}} }{{2{\left\| {{\rm {\bf
u}}} \right\|}}}}\varepsilon _{ijkl} u^{j}S^{kl}{\rm .}
\end{equation}
Умова Матісона-Пірані (\ref{matsyuk:4}) дозволяє обернути формулу
(\ref{matsyuk:17}):
\begin{equation}\label{matsyuk:18}
S_{ij} =
{\dfrac{\sqrt {{\left| {g} \right|}}} {{\left\| {{\rm
{\bf u}}} \right\|}}}
\varepsilon _{ijkl} u^{k}  {{s}} ^{l}{\rm .}
\end{equation}
Тепер ми готові подати систему рівнянь Діксона-Матісона
(\ref{matsyuk:1},~\ref{matsyuk:2},~\ref{matsyuk:4}) на
закріпленій формулою (\ref{matsyuk:11}) масовій оболонці в иньшому
рівнозначному вигляді,
куди входитимуть похідні від координат частки до третього порядку включно.
Еквівалентність порушиться тільки в тім випадку, коли ми забудемо про
вихідну умову Матісона-Пірані (\ref{matsyuk:4}). Розгляньмо систему рівнянь:
\begin{gather}\label{matsyuk:19}
\begin{split}
\varepsilon _{ijkl} \ddot {u}^{j}u^{k}  {{s}} ^{l} -
3\,{\frac{{{\rm {\bf \dot {u}}} \cdot {\rm {\bf u}}}}{{{\rm {\bf
u}}^{2}}}}\,\varepsilon _{ijkl} \dot {u}^{j}u^{k}  {{s}} ^{l} +
{\frac{{m}}{{\sqrt {{\left| {g} \right|}} }}}{\left[ {\left( {{\rm
{\bf \dot {u}}} \cdot {\rm {\bf u}}} \right)u_{i} - {\rm {\bf
u}}^{2}\dot {u}_{i}}  \right]} \\ = {\frac{{{\rm {\bf
u}}^{2}}}{{2}}}R_{ij}{}^{mn} \varepsilon _{mnkl} u^{j}u^{k}  {{s}}
^{l}
\end{split}
 \\
u^{2}{\rm {\bf \dot {  {{s}}} }} + \left( {{\rm {\bf   {{s}}} } \cdot {\rm
{\bf \dot {u}}}} \right)\;{\rm {\bf u}} = 0{\rm .}
\label{matsyuk:20}
\\
{\rm {\bf   {{s}}} } \cdot {\rm {\bf u}} = 0{\rm .}
\label{matsyuk:21}
\end{gather}

\textbf{Твердження}\textit{. В області} ${\rm {\bf u}}^{2} \ne 0$\textit{ система рівнянь
(\ref{matsyuk:19},~\ref{matsyuk:20}, \ref{matsyuk:21}, \ref{matsyuk:18},
(\ref{matsyuk:18})}$^{^{\updot}}$\textit{)
алґебрично рівнозначна
системі
(\ref{matsyuk:16},~\ref{matsyuk:9}, \ref{matsyuk:4},
\ref{matsyuk:11}, \ref{matsyuk:17},~(\ref{matsyuk:17})$^{^{\updot}}$).}

\textbf{\textit{Доведення}}\textit{.} З метою отримати рівняння (\ref{matsyuk:18}),
домножимо
означення (\ref{matsyuk:17}) зовнішнім чином на вектор ${\rm {\bf u}}$,
перейдемо до
двоїстого рівняння і використаємо умову Матісона-Пірані (\ref{matsyuk:4}).
Навпаки,
домноживши (\ref{matsyuk:18}) зовнішнім чином на вектор ${\rm {\bf u}}$
і перейшовши до
двоїстого рівняння, одержимо, з використанням умови (\ref{matsyuk:21}),
рівняння (\ref{matsyuk:17}).

Рівняння (\ref{matsyuk:19}) і (\ref{matsyuk:16}) рівнозначні з огляду на
(\ref{matsyuk:18}).

Рівняння (\ref{matsyuk:9}) записуємо у вигляді

\[
\left( {\sqrt {{\left| {g} \right|}} \,\varepsilon _{ijkl} u^{j}S^{kl}}
\right)^{^{\updot}}  - {\frac{{\sqrt {{\left| {g} \right|}}
}}{{g}}}e^{ijkl}\dot {u}_{j} S_{kl} = 0{\rm ,}
\]
де $\frac{\left( \sqrt {\left| g \right|}  \right)^2}
{g} = - 1$. У перший доданок підставляємо формулу (\ref{matsyuk:17}),
а у другий доданок
-- формулу (\ref{matsyuk:18}) і одержуємо подвоєне рівняння (\ref{matsyuk:20}).

Зворотній шлях довший. Спочатку домножуємо (\ref{matsyuk:20}) зовнішнім чином на вектор
$2{\rm {\bf u}}$ і переходимо до двоїстого рівняння

\[
{\rm {\bf u}}^{2}\left( {\sqrt {{\left| {g} \right|}} \,\varepsilon _{ijkl}
u^{k}  {{s}} ^{l}} \right)^{^{\updot} } - {\rm {\bf u}}^{2}\sqrt {{\left| {g}
\right|}} \,\varepsilon _{ijkl} \dot {u}^{k}  {{s}} ^{l} = 0{\rm .}
\]

У перший доданок підставляємо формулу (\ref{matsyuk:18}), а у другий
доданок -- формулу
(\ref{matsyuk:17}). Одержимо:

\[
{\left\| {{\rm {\bf u}}} \right\|}^{3}\dot {S}_{ij} + {\left\| {{\rm {\bf
u}}} \right\|}\,\left( {u_{j} S_{ik} \dot {u}^{k} - u_{i} S_{jk} \dot
{u}^{k}} \right) = 0{\rm .}
\]

Знову домножуючи зовнішнім чином на вектор ${\frac{{2}}{{{\left| {g}
\right|}}}}{\rm {\bf u}}$ і перейшовши до двоїстого виразу, отримаємо
(\ref{matsyuk:9}).$\blacksquare$

Замість рівняння (\ref{matsyuk:20}) в систему
(\ref{matsyuk:19},~\ref{matsyuk:20},~\ref{matsyuk:21}) можна вписати таке:
\begin{equation}\label{matsyuk:22}
{\rm {\bf \dot {  {{s}}} }} = {\frac{{\sqrt {{\left| {g} \right|}}
}}{{2m\,{\rm {\bf u}}^{2}}}}\varepsilon _{mnkl} R_{ij}{}^{mn} u^{j}u^{k}  {{s}}
^{l}  {{s}} ^{i}\,{\rm {\bf u}}{\rm .}
\end{equation}

У справедливості вказаної заміни переконаємось, згорнувши рівняння
(\ref{matsyuk:19}) з
вектором ${\rm {\bf   {{s}}} }$ і використавши (\ref{matsyuk:21}).

В пласк\'ому просторі-часі вертун ${\rm {\bf   {{s}}} }$ зберігається весь,
як бачимо з рівняння (\ref{matsyuk:22}).

\section{Варіяційне узагальнення рівнянь Матісона\\без ґравітації}\label{sec:mylabel2}
Зробімо деякі алґебричні перетворення над рівнянням
(\ref{matsyuk:19}). Для спрощення
запису використовуватимемо загальноприйняті позначення скісної векторної
алґебри, а також позначимо праву частину рівняння
(\ref{matsyuk:19}) літерою ${\rm {\bf F}}$. Згортка рівняння
(\ref{matsyuk:19}) з чотири-вектором ${\rm {\bf   {{s}}} }$ одразу
приводить до співвідношення
\begin{equation}\label{matsyuk:23}
m\,\left( {{\rm {\bf   {{s}}} } \wedge {\rm {\bf u}}} \right) \cdot \left(
{{\rm {\bf \dot {u}}} \wedge {\rm {\bf u}}} \right) = - \,\left( {{\rm {\bf
F}} \cdot {\rm {\bf   {{s}}} }} \right){\rm ,}
\end{equation}
де під величиною $\left( {{\rm {\bf   {{s}}} } \wedge {\rm {\bf u}}} \right)
\cdot \left( {{\rm {\bf \dot {u}}} \wedge {\rm {\bf u}}} \right)$ розуміємо
скалярний добуток бівекторів,

\[
\left( {{\rm {\bf   {{s}}} } \wedge {\rm {\bf u}}} \right) \cdot \left( {{\rm
{\bf \dot {u}}} \wedge {\rm {\bf u}}} \right) = {\rm {\bf u}}^{2}\left(
{{\rm {\bf   {{s}}} } \cdot {\rm {\bf \dot {u}}}} \right) - \left( {{\rm
{\bf   {{s}}} }
\cdot {\rm {\bf u}}} \right)\,\left( {{\rm {\bf u}} \cdot {\rm
{\bf \dot {u}}}} \right){\rm .}
\]

Оперуючи співвідношенням (\ref{matsyuk:23}), легко переконатися у справедливості ще й
такого співвідношення, незалежно від умови (\ref{matsyuk:21}):
\begin{equation}\label{matsyuk:24}
{\frac{{\left( {{\rm {\bf u}} \cdot {\rm {\bf \dot {u}}}} \right)}}{{{\rm
{\bf u}}^{2}}}} = {\frac{{\left( {{\rm {\bf   {{s}}} } \wedge {\rm {\bf u}}}
\right) \cdot \left( {{\rm {\bf   {{s}}} } \wedge {\rm {\bf \dot {u}}}}
\right)}}{{\left( {{\rm {\bf   {{s}}} } \wedge {\rm {\bf u}}} \right)^{2}}}}
- {\frac{{{\rm {\bf   {{s}}} } \cdot {\rm {\bf u}}}}{{{\rm {\bf u}}^{2}\left(
{{\rm {\bf   {{s}}} } \wedge {\rm {\bf u}}} \right)^{2}}}}{\frac{{\,\left(
{{\rm {\bf F}} \cdot {\rm {\bf   {{s}}} }} \right)}}{{m}}}{\rm .}
\end{equation}

Поділімо тепер рівняння (\ref{matsyuk:19}) на величину ${\left\| {{\rm {\bf   {{s}}} }
\wedge {\rm {\bf u}}} \right\|}^{3}$ та скористаймо з наслідку (\ref{matsyuk:24}).
Одержимо таке

\textbf{Твердження}. \textit{На поверхні (\ref{matsyuk:21}) рівняння
(\ref{matsyuk:19}) є алґебрично рівнозначним до такого:}
{\setlength{\multlinegap}{0pt}
\begin{multline}
\label{matsyuk:25}
{\frac{{\varepsilon _{ijkl} \ddot {u}^{j}u^{k}  {{s}} ^{l}}}{{{\left\| {{\rm
{\bf   {{s}}} } \wedge {\rm {\bf u}}} \right\|}^{3}}}} - 3\,{\frac{{\left(
{{\rm {\bf   {{s}}} } \wedge {\rm {\bf u}}} \right) \cdot \left( {{\rm
{\bf   {{s}}} }
\wedge {\rm {\bf \dot {u}}}} \right)}}{{{\left\| {{\rm {\bf   {{s}}
}} \wedge {\rm {\bf u}}} \right\|}^{5}}}}\varepsilon _{ijkl} \dot
{u}^{j}u^{k}  {{s}} ^{l} - {\frac{{m}}{{{\left\| {{\rm {\bf   {{s}}} } \wedge
{\rm {\bf u}}} \right\|}^{3}}}}\left( {{\rm {\bf u}}^{2}\dot {u}_{i} -
\left( {{\rm {\bf u}} \cdot {\rm {\bf \dot {u}}}} \right)\;u_{i}}  \right)
=
{\frac{{F_{i}} }{{{\left\| {{\rm {\bf   {{s}}} } \wedge {\rm {\bf u}}}
\right\|}^{3}}}}
\end{multline}}

Розглянемо рівняння (\ref{matsyuk:25}) при відсутності ґравітації.
Знаком „$ * $“
позначимо операцію переходу до двоїстого тензора.

\textbf{Твердження}\textit{. Рівняння}
\begin{equation}\label{matsyuk:26}
{\frac{{ *\, {\rm {\bf \ddot {u}}} \wedge {\rm {\bf u}} \wedge {\rm
{\bf   {{s}}} }}}{{{\left\| {{\rm {\bf   {{s}}} } \wedge {\rm {\bf u}}}
\right\|}^{3}}}} - 3\,{\frac{{\left( {{\rm {\bf   {{s}}} } \wedge {\rm {\bf
u}}} \right) \cdot \left( {{\rm {\bf   {{s}}} } \wedge {\rm {\bf \dot {u}}}}
\right)}}{{{\left\| {{\rm {\bf   {{s}}} } \wedge {\rm {\bf u}}}
\right\|}^{5}}}}\; * {\rm {\bf \dot {u}}} \wedge {\rm {\bf u}} \wedge {\rm
{\bf   {{s}}} } + {\frac{{\mu} }{{{\left\| {{\rm {\bf   {{s}}} }}
\right\|}^{3}{\left\| {{\rm {\bf u}}} \right\|}^{3}}}}\left( {{\rm {\bf
u}}^{2}{\rm {\bf \dot {u}}} - \left( {{\rm {\bf u}} \cdot {\rm {\bf \dot
{u}}}} \right)\,{\rm {\bf u}}} \right) = 0
\end{equation}
\textit{при постійних} $\mu$  \textit{та} ${\rm {\bf   {{s}}} }$\textit{ має
варіяційне походження. Воно описує рух вільної буцім-клясичної
дзиґи з масою (\ref{matsyuk:7}). Величина (\ref{matsyuk:6}) є першим його
інтеґралом.}

Рівняння (\ref{matsyuk:26}) можна отримати від кожної з цілого сімейства функцій Ляґранжа.
Нехай ${\left\{ {{\rm {\bf e}}_{\left( {k} \right)}}  \right\}}_{k =
\overline {0,\,3}}  $ значить псевдоортонормовану базу в пласк\'ому
просторі-часі. Нехай
\begin{equation}\label{matsyuk:27}
 L_{\left( {k} \right)} = {\frac{{ *\, {\rm {\bf \dot {u}}} \wedge {\rm {\bf
u}} \wedge {\rm {\bf   {{s}}} } \wedge {\rm {\bf e}}_{\left( {k} \right)}
}}{{\left( {u_{k} {\rm {\bf   {{s}}} } -   {{s}} _{k} {\rm {\bf u}}}
\right)^{2} - \left( {{\rm {\bf   {{s}}} } \wedge {\rm {\bf u}}}
\right)^{2}}}}\left( {{\rm {\bf   {{s}}} }^{2}u_{k} + \left( {{\rm {\bf   {{s}}} }
\cdot {\rm {\bf u}}} \right)\,  {{s}} _{k}}  \right){\rm ,}\quad k =
\overline {0,\,3} .
\end{equation}

Кожна з функцій Ляґранжа
\begin{equation}\label{matsyuk:28}
{\frac{{\mu} }{{{\left\| {{\rm {\bf   {{s}}} }} \right\|}^{3}}}}{\left\|
{{\rm {\bf u}}} \right\|} + {\frac{{L_{\left( {k} \right)}} }
{{\left\|{\rm {\bf   {{s}}} }\right\|^{2}{\left\| {{\rm {\bf   {{s}}} } \wedge {\rm {\bf u}}}
\right\|}}}}
\end{equation}
породжує рівняння (\ref{matsyuk:26}) (з постійним чотири-вектором
${\rm {\bf   {{s}}} }$,
який не підлягає варіяції).

Ми не знаємо, чи існують Лоренц-інваріянтні та глобально означені функції
Ляґранжа для рівняння (\ref{matsyuk:26}). Вираз у лівій частині рівняння
(\ref{matsyuk:26}) теж не є
інваріянтом. Сам\'е рівняння (\ref{matsyuk:26}), одначе, наділене
лоренцівською симетрією в
тому розумінні, яке прийняте у теорії диференційних рівнянь. Це
забезпечується векторним характером лівої частини рівняння
(\ref{matsyuk:26}). Якщо

\[
X = \Omega ^{kl}u_{k} {\frac{{\partial} }{{\partial u^{l}}}} + \Omega
^{kl}\dot {u}_{k} {\frac{{\partial} }{{\partial \dot {u}^{l}}}} + \Omega
^{kl}\ddot {u}_{k} {\frac{{\partial} }{{\partial \ddot {u}^{l}}}} + \Omega
^{kl}  {{s}} _{k} {\frac{{\partial} }{{\partial   {{s}} ^{l}}}}
\]
є ґенератором псевдоортогональних перетворень, а ${\rm {\bf v}}$ --
довільний вектор, тоді

\[
Xv^{i} = - \eta ^{ij}\Omega _{jk} v^{k}{\rm ,}
\]
де $\eta ^{ij}$ -- це є постійна діяґональна матриця канонічної контраваріянтної метрики
псевдоевклідівського простору. В застосуванні до рівняння (\ref{matsyuk:26})
бачимо, що
множина світових ліній -- розв'язків рівняння (\ref{matsyuk:26}) не змінюється
під дією
лоренцівських перетворень. При цій нагоді згадаємо, що лоренц-інваріянтних
варіяційних рівнянь третього порядку, які б не містили жодних сторонніх
векторних чи тензорних параметрів, не існує зовсім~\cite{matsyuk:2}.

Пригадавши означення варіяційного імпульсу, ${\rm {\bf p}} =
{\frac{{\partial L}}{{\partial {\rm {\bf u}}}}} - \left( {{\frac{{\partial
L}}{{\partial {\rm {\bf \dot {u}}}}}}} \right)^{ \updot} $ та вираз для
рівняння Ойлера-Пуасона, $ - \left( {{\frac{{\partial L}}{{\partial {\rm
{\bf u}}}}}} \right)^{ \updot}  + \left( {{\frac{{\partial L}}{{\partial {\rm
{\bf \dot {u}}}}}}} \right)^{ \updot \updot}  = 0$, можемо з рівняння
(\ref{matsyuk:26})
легко узріти вираз для кількости руху частки (вибір знаку реґулюється
знаками доданків у формулі (\ref{matsyuk:28})):
\begin{equation}\label{matsyuk:29}
 - {\rm {\bf p}} = {\frac{{ *\, {\rm {\bf \dot {u}}} \wedge {\rm {\bf u}}
\wedge {\rm {\bf   {{s}}} }}}{{{\left\| {{\rm {\bf   {{s}}} } \wedge {\rm {\bf
u}}} \right\|}^{3}}}} + {\frac{{\mu} }{{{\left\| {{\rm {\bf   {{s}}} }}
\right\|}^{3}{\left\| {{\rm {\bf u}}} \right\|}}}}{\rm {\bf u}}{\rm .}
\end{equation}

Порівняємо поміж собою вирази для кількости руху
(\ref{matsyuk:29}) і (\ref{matsyuk:13}), скориставши з
означення (\ref{matsyuk:18}) для величини $S_{ij} $.
Узгодивши масу частки співвідношенням
(\ref{matsyuk:7}) та використовуючи означення (\ref{matsyuk:11}), одержимо:

\[
{\rm {\bf P}} = - {\frac{{{\left\| {{\rm {\bf   {{s}}} } \wedge {\rm {\bf
u}}} \right\|}^{3}}}{{{\left\| {{\rm {\bf u}}} \right\|}^{3}}}}{\rm {\bf
p}}{\rm .}
\]
Множник пропорційности є сталим в пласк\'ому просторі-часі з огляду
на збереження інтеґралу руху (\ref{matsyuk:6}) і постійність $4$-вектора
${\rm {\bf   {{s}}}}$.

{\textit{Зауваження.}} В одній з попередніх робіт~\cite{matsyuk:8} ми вказали
на спосіб отримати рівняння~(\ref{matsyuk:26}) із загальної засади
варіяційности в поєднанні з вимогою лоренцівської симетрії.

\section{Тривимірний пласк\'ий простір-час}\label{sec:mylabel3}
Система рівнянь
(\ref{matsyuk:19},~\ref{matsyuk:20},~\ref{matsyuk:21})
при відсутності ґравітації допускає рух у
площині, коли $u^{3} = \dot {u}^{3} = \ddot {u}^{3} = 0$. У три-вимірному
пласк\'ому просторі-часі рівняння (\ref{matsyuk:19}) прибере
форми (гляди~\cite{matsyuk:1}):
\begin{equation}\label{matsyuk:30}
  {{s}} ^{3}\left( {{\frac{{{\rm {\bf \ddot {u}}}\times {\rm {\bf
u}}}}{{{\left\| {{\rm {\bf u}}} \right\|}^{3}}}} - 3{\frac{{{\rm {\bf \dot
{u}}} \cdot {\rm {\bf u}}}}{{{\left\| {{\rm {\bf u}}} \right\|}^{5}}}}{\rm
{\bf \dot {u}}}\times {\rm {\bf u}}} \right) + {\frac{{m}}{{{\left\| {{\rm
{\bf u}}} \right\|}^{3}}}}\left( {{\rm {\bf u}}^{2}{\rm {\bf \dot {u}}} -
\left( {{\rm {\bf \dot {u}}} \cdot {\rm {\bf u}}} \right)\;{\rm {\bf u}}}
\right) = 0{\rm .}
\end{equation}

В роботах \cite{matsyuk:2} і \cite{matsyuk:7} доведено, що рівняння (\ref{matsyuk:30}) з довільними постійними параметрами
$  {{s}} ^{3}=\eta ^{33}  {{s}}_3$ та $m$ -- це єдине релятивіське рівняння третього порядку, яке
ще може мати варіяційне походження. Відомі нам функції Ляґранжа для рівняння
(\ref{matsyuk:30}), як і в попередньому випадку, утворюють ціле сімейство, параметризоване
вибором псевдоевклідівської 3-бази ${\left\{ {{\rm {\bf e}}_{\left( {k}
\right)}}  \right\}}_{k = \overline {0,\,2}}  $ в просторі-часі:

\[
 L_{\left( k \right)} =   {{s}} ^{3}{\frac{{u^{k}{\left[ {{\rm {\bf \dot {u}}},{\rm {\bf
u}},{\rm {\bf e}}_{\left( {{ { k}}} \right)}}  \right]}}}{{{\left\|
{{\rm {\bf u}}} \right\|}\;{\left\| {{\rm {\bf u}}\times {\rm {\bf
e}}_{\left( {{ { k}}} \right)}}  \right\|}^{2}}}} + m{\left\| {{\rm
{\bf u}}} \right\|}{\rm ,}\quad k = \overline {0,\,2} .
\]

Якщо компоненти $  {{s}} ^{1}$ і $  {{s}} ^{2}$ відсутні, то з формули
(\ref{matsyuk:21})
видно, що і $  {{s}} ^{0} = 0$, так що ${\left\| {{\rm {\bf   {{s}}} }}
\right\|} =   {{s}} ^{3}$. Вираз для величини кількости руху частки і в цьому
випадку добре проглядається у лівій частині рівняння
(\ref{matsyuk:30}) (гляди ще~\cite{matsyuk:1}). Він співпадає із
тим виразом, який пізніше розглядався в роботі~\cite{matsyuk:5}:

\[
 - {\rm {\bf p}} = {\frac{{{\rm {\bf \dot {u}}}\times {\rm {\bf
u}}}}{{{\left\| {{\rm {\bf u}}} \right\|}^{3}}}} + {\frac{{m}}{{{\left\|
{{\rm {\bf   {{s}}} }} \right\|}}}}{\frac{{{\rm {\bf u}}}}{{{\left\| {{\rm
{\bf u}}} \right\|}}}}{\rm .}
\]

\begin{center}
{\bf VARIATIONAL GENERALIZATION \\ OF FREE RELATIVISTIC TOP }

\bigskip
\href{http://iapmm.lviv.ua/12/eng/files/st_files/matsyuk.htm}{\em Roman MATSYUK}

\bigskip
Pidstryhach Institute for Applied Problems\\ in Mechanics and
Mathematics,\\ Ukrainian National Academy of Sciences,  \\3-b
Naukova Str., Lviv 79060
\end{center}
We prove that well known first-order (in spin, momentum, and
space-time coordinates) equations of motion of relativistic top
are equivalent to the third-order equations of Mathisson on the
surface of the Mathisson-Pirani auxiliary constraint. We then
consider these third-order equations in flat space-time with
constant spin $4$-vector and invent a Lagrange function for them.
Allowing physical interpretation to be applied to the complete set
of extremals yields a whole spectrum of spin-dependent effective
``proper mass'' of the relativistic top.
\end{document}